\documentclass[preprint,superscriptaddress,amssymb,amsmath]{revtex4-1}

\usepackage{graphicx}% Include figure files
\usepackage{dcolumn}% Align table columns on decimal point
\usepackage{bm}% bold math
\usepackage{bm, color}

\begin{document}

\title{Comprehensive scan for nonmagnetic Weyl semimetals with nonlinear optical response}% Force line breaks with \\

\author{Qiunan Xu}
\thanks{These two authors contributed equally}
\affiliation{Max Planck Institute for Chemical Physics of Solids, 01187 Dresden, Germany}
\author{Yang Zhang} 
\thanks{These two authors contributed equally}
\affiliation{Max Planck Institute for Chemical Physics of Solids, 01187 Dresden, Germany}
\affiliation{Leibniz Institute for Solid State and Materials Research, 01069 Dresden, Germany}
\author{Klaus Koepernik}
\affiliation{Leibniz Institute for Solid State and Materials Research, 01069 Dresden, Germany}
\author{Wujun Shi}
\affiliation{Max Planck Institute for Chemical Physics of Solids, 01187 Dresden, Germany}
\affiliation{School of Physical Science and Technology, ShanghaiTech University, Shanghai 200031, China}
\author{Jeroen van den Brink}
\affiliation{Leibniz Institute for Solid State and Materials Research, 01069 Dresden, Germany}
\author{Claudia Felser}
\affiliation{Max Planck Institute for Chemical Physics of Solids, 01187 Dresden, Germany}
\affiliation{Center for Nanoscale Systems,Faculty of Arts and Sciences, Harvard University,11 Oxford Street, LISE 308 Cambridge, MA 02138, USA}
\author{Yan Sun}
\email{Yan.Sun@cpfs.mpg.de}
\affiliation{Max Planck Institute for Chemical Physics of Solids, 01187 Dresden, Germany}

\begin{abstract}
As the development of topological band theory, comprehensive databases 
about time reversal and crystalline symmetries protected nonmagnetic 
topological materials were developed via first-principles calculations 
recently. However, owing to the low symmetry requirement of Weyl points, 
the symmetry-based topological indicator cannot be applied to Weyl 
semimetals (WSMs). Hitherto, the WSMs with Weyl points in arbitrary 
positions are still absent in the well-known databases. 
In this work, we develop an efficient algorithm to search for Weyl points automatically
and establish a database of nonmagnetic WSMs with Weyl points near Fermi level 
based on the total experimental noncentrosymmetric crystal structures 
in the Inorganic Crystal Structure Database (ICSD).
Totally 46 Weyl semimetals were discovered to have 
nearly clean Fermi surface and Weyl points near Fermi level within 300 meV,
and 9 of them are chiral structures which may host the quantized circular photogalvanic effect. 
In addition, the nonlinear optical response is studied and giant shift current is explored in the end. 
Besides nonmagnetic WSMs, our powerful tools can also be used in the discovery of magnetic topological materials.
\end{abstract}

\maketitle

%Background and introduction
\section*{Introduction}

Since the discovery of topological insulators~\cite{Kane2005,bernevig2006quantum,konig2007quantum,fu2007topological,hsieh2008topological,Hasan2010ku,Qi2011RMP},
topological band theory has been widely applied to solid state materials~\cite{chiu2016classification,po2017symmetry,song2018quantitative,song2017diagnosis}
and leads to the further classification of both insulators~\cite{kitaev2009periodic,fu2011topological,hsieh2012topological,RJ2,ando2015topological,RJ1,schindler2018higher}
and semimetals~\cite{Wan2011,Burkov2011de,Wang2012,Young2012,Weng2015,huang2015weyl,
Xu2015TaAs,Lv2015TaAs,bradlyn2016beyond,bradlyn2017topological,armitage2018weyl}
based on the quantum topological invariants. 
The development of the topological semimetals stimulated extensive research in understanding a variety of topological properties.
Very recently, some comprehensive databases for the symmetry protected nonmagnetic topological states
were established by three independent works~\cite{zhang2019catalogue,tang2019comprehensive,vergniory2019complete}.
These useful databases contain not only the topological insulators, but also the semimetals with degeneracy points located at high-symmetric lines or points.
However, in these databases, an important category of topological materials, eg. the Weyl semimetal (WSM)
whose degeneracy point can locate at any position in the entire Brillouin zone, is still absent.

Generally speaking, topological semimetals are the systems whose conduction and valence bands cross with each other in the Brillouin zone 
and cannot be removed by perturbations of Hamiltonian without breaking specific symmetries.
Topological WSM is such a material with double degeneracy Weyl points.
Compared with other topological states, 
the WSM is extraordinary since the topology is not protected by any crystalline symmetry.
It can only appear in the system whose spin degeneracy is removed by breaking time reversal or spacial inversion symmetry.
In this case, the bands that create the Weyl points take the $2\times2$ Weyl equation form, $H=\pm v_{F}\vec{k}\cdot\vec{\sigma}$, 
which can be viewed as half of the massless Dirac equation with defined chirality in three-dimensional space~\cite{weyl1929elektron}.
Because a two-bands model Hamiltonian can be described by a linear combination of three Pauli matrices generally, 
the linear band crossing can be obtained by tuning the parameters without considering any additional conditions.
Therefore, besides time reversal or spacial inversion symmetry, the lattice periodic boundary condition is the unique requirement for Weyl points. 
Accordingly, the identification of topology based on the characteristics of high-symmetry points does not work anymore for WSMs,
and that is why the WSMs are still absent in the comprehensive databases~\cite{zhang2019catalogue,tang2019comprehensive,vergniory2019complete}.

In this work, we only focus on the experimental structures to
complete the databases by including nonmagnetic WSMs.
The total information about their topological properties is listed in the Supplementary Information. 
At the same time, two effective tools used for generating Wannier 
function automatically and searching Weyl points are developed in Fig. 1. 
The accuracy of them is essential for finding materials with observable properties of Weyl points, 
Some of them have already been theoretically predicted and experimentally confirmed, such as
the family of TaAs~\cite{Weng2015,huang2015weyl,Xu2015TaAs,Lv2015TaAs}.
Besides type-I and type-II WSMs, 9 chiral WSMs with inversion and 
mirror symmetries breaking are also listed in our database. 
For chiral WSMs, the truly quantized circular photogalvanic effect (CPGE)~\cite{de2017quantized,RhSi_hasan,flicker2018chiral,chiral_hasan}
could be induced by the energy difference of Weyl points with opposite chirality. 
Moreover, we explored the bulk photovoltaic effect (BPVE)~\cite{design_SC} of all these noncentrosymmetric WSMs. 
A giant BPVE which reach up to $700 \mu A/V^2$ is predicted, 
and some of them can even extend to the regime of visible light.

\section*{Results}
Our automatic workflow for searching nonmagnetic WSMs is shown in Fig. 1.
It includes two crucial algorithms, the Wannier function generation and Weyl points search.
Constrained by the accuracy of the density functional theory (DFT), we don't consider 
the systems with strongly correlated f-elections, namely Lanthanide and Actinide series except element La,
while beginning our work with the experimental noncentrosymmetric crystal structures in the Inorganic Crystal Structure Database (ICSD)~\cite{Mariette2004}.
Here we remain some systems with 3d-elections, which may also be strongly correlated systems, 
to make sure that we don't omit some potential candidates.
After removing alloys and repeated structures, totally 8896 compounds left. 
Among them, all the systems with the total magnetic moments exceeding 0.05 $\mu_{B}$ per unit cell 
and magnetic elements V, Mn, Fe, Co and Ni are regarded as magnetic compounds.
Furthermore, the gap and the charge carrier density ($N$) of the nonmagnetic phases are checked. 
Because we are only interested in WSMs with observable topological Weyl-points-related properties, 
the insulators and good metals (with $N>2\times10^{22}/cm^{3}$, which is obtained by the comparison with Co$_2$VGa~\cite{manna2018colossal}) 
will not be considered in the further calculations.

After all the screening processes above, we obtain the list of the nonmagnetic noncentrosymmetric 
experimental crystal structures who may possess Weyl points near Fermi level.
For these compounds, the atomic-orbital-like Wannier functions are constructed by an effective algorithm automatically
as shown in Fig. 1{\bf b}.
The crucial problem to sole for this step is the setting of the fitting windows with appropriate projection orbital basis sets.
In the nonmagnetic first-principles calculations, we select $s$ orbitals of alkaline-earth metals, 
$d$ orbitals of transition metals and $s+p$ orbitals of $p$-block elements as the minimum basis sets. 
Thus, we can label the maximum weight value among the selected band as $r$. 
The weight represent the ratio of selected basis projection.
If the weight $r$ is smaller than 0.8, the minimum basis sets will be insufficient to produce the DFT band structure; 
hence, we switch the bases to the maximum basis sets which include $s+p+d$ orbitals of all the elements. 
With the selected basis sets, the inner projection energy window can be easily fixed as 90\% of the maximum weight value $r$. 
Subsequently, we loop over the outer projection energy window and check the mean absolute error
between the DFT and Wannier band structures around the Fermi level until the error is smaller than 0.02 eV.
Meanwhile, it is noteworthy that a good projection can be always guaranteed 
because of the large band gap between valence and core states.

Using the high-symmetry tight-binding (TB) model Hamiltonians obtained in the construction of Wannier functions,
the Weyl points can be scanned. 
Since the existence of Weyl points in noncentrosymmetric nonmagnetic system only requires the translational symmetry,
the fact that the Weyl points would locate at generic k-points without any special symmetry renders the search difficult. 
The typical way to search for the Weyl points use the band gap as a criterion. 
However, this method is much easy to reach the local minimum value, thereby losing some Weyl points. 
In this work, we use the Berry phase of the parallelepiped corresponding to the k-mesh of the reciprocal space as the criterion.
Initially, based on the TB model Hamiltonian,
we generate a relatively sparse $51\times51\times51$ k-mesh along the three reciprocal vectors 
to calculate the Berry phase of each parallelepiped. 
If the value is larger than $0.01\times2\pi$, Weyl points might be present inside the parallelepiped.
For this type of parallelepiped, we increase the k-mesh finer and repeat calculating Berry phase. 
Eventually, the Berry phase will increase in some fine parallelepiped and finally reach a value
$n\times2\pi$, where $n$ is an integer, corresponding to a single Weyl point. 
In our calculations, the parallelepiped is refined to around 1/1600$^3$ of the reciprocal lattice 
to obtain the Berry phase with $n\geqslant0.99$.

After searching the Weyl points, we have a raw list of the WSMs with all the information of Weyl points.
It may also have some magnetic materials such as antiferromagnets. 
Thus, we process a double check here to modify the raw list,
and the list of nonmagnetic WSMs are obtained.

Despite having as many as 8896 noncentrosymmetric candidates,
we only find 46 nonmagnetic WSMs with Weyl points near the Fermi level within 300 meV, 
including 9 chiral crystal structures.
This result express that the good nonmagnetic WSMs are difficult to find. 
The details of every WSMs that we find by our algorithm are given in Supplementary Table 1. 
%Depending on the specific situation, the WSMs can be further classified.
Until now, the Weyl points have been classified into type-I and type-II according to the shape of the linear band crossings. 
The Fermi surface would shrink to a singularity point for the type-I Weyl point,
while the type-II Weyl point presents a Fermi surface with a linear crossing which is composed of electron and hole~\cite{soluyanov2015type}. 
On the other hand, another main difference is the precondition of chiral anomaly.
It can exist along any direction in the type-I WSM as long as electric and magnetic fields are not perpendicular to each other,
but only exist when the electric and magnetic fields point to some selected direction in the type-II WSM.

\section*{Discussion}

\subsection*{Representative Materials}
Here, we choose two typical candidates covering each class of WSM as examples. 
For type-I WSM, we highlight an orthorhombic sulphide TaAgS$_3$ with space group $Cmc2_1$ (No. 36) as shown in Fig. 2{\bf a}.
The energy dispersion calculated by DFT is given in Fig. 2{\bf d}.
It presents a semimetallic structure with small electron and hole pockets located around $\Gamma$ and $Y$ points, respectively.
With inversion symmetry breaking, the spin degeneracy is lifted.
Finally, two pairs of Weyl points, a minimum number of Weyl points in time-reversal-symmetric systems, would be generated.
Because of the $C_{2z}$ rotation and the time reversal symmetry, the Weyl points are constrained in the $k_z=0$ plane as shown in Fig. 2{\bf b}.
All the Weyl points are related with each other and located at the same energy,
only approximately 30 meV below the charge neutral point, owing to the existence of the mirror planes $m_x$ and $m_y$, 
On the other hand, low charge carrier density can make the Weyl points observable by transport measurements. 
The Weyl points are further confirmed by the (001) surface state calculation. 
When these Weyl points are projected into (001) surface, the Weyl points with different chirality are projected into different position.
If the chemical potential is set at the energy of the Weyl points, the Fermi arcs connecting the Weyl points can be observed as shown in Fig. 2{\bf e}.
The Fermi arc length can be $\sim15\%$ of the reciprocal lattice vector, which makes it easy to detect by surface detection methods, 
such as angle-resolved photoemission spectroscopy (ARPES) and scanning tunnelling microscopy (STM).

The other example Ag$_2$S is a hybrid WSM~\cite{Hybrid} containing both two kinds of Weyl points. 
It also has an orthorhombic crystal structure with space group $P2_12_12_1$ (No. 19), as shown in Fig. 3{\bf a}. 
Different from TaAgS$_3$, without considering spin-orbital coupling (SOC), the band inversion between the $p$ orbital of S and the $d$ orbital of Ag forms two Dirac points 
on the high-symmetric lines $\Gamma-X$ and $\Gamma-Y$ which is protected by the $C_{2}$ rotation 
symmetry~\cite{zhang2019catalogue,tang2019comprehensive,vergniory2019complete}. 
Furthermore, turning on SOC leads to the separation of the Dirac points and 4 pairs of Weyl nodes appear, 
since SOC breaks the spin rotation symmetry.
The locations of Weyl points in the Brillouin zone shown in Fig. 3{\bf b} connected with each other by time reversal, $C_{2x}$ and $C_{2y}$ rotation symmetries.
Through the band structure along the k-path crossing one pair of Weyl points in Fig. 3{\bf d}, 
we find these Weyl points belong to two different classes, the type-I Weyl points with chirality +1 (W1) and the type-II Weyl points with chirality -1 (W2).
They are distinguished by different color in Fig. 3.
At the same time, the type-II Weyl points W2 can also be located by the linear-crossing-like Fermi surface in Fig. 3{\bf e} if we put the chemical potential at W2.

One point we need to emphasized is that Ag$_2$S does not present any mirror symmetry.
Although the mirror symmetry is not necessary for the existence of Weyl points, 
the chirality of Weyl points must be odd with mirror operations.
Therefore, the mirror symmetry forces one pair of Weyl points with opposite chirality locate at the same energy. 
In another words, WSM without mirror symmetry exhibits a net chirality if the Fermi energy is 
located between one pair of Weyl points with opposite chirality. 
One of the most important properties for mirror-symmetry-broken WSMs is the quantized CPGE,
which is the only possible quantized signal induced by the monopole Weyl point so far.
This can also be applied to Ag$_2$S with W1 $\sim23$ meV above the charge neutral point and W2 
$\sim28$ meV below the charge neutral point, as illustrated in Fig. 3{\bf d}. 
Because the charge carrier density is only around $5\times10^{20}/cm^{3}$ and 
the contribution of the trivial Fermi surfaces can be neglected, 
Ag$_2$S may be an ideal candidate to observe the quantized CPGE.

The chiral WSM Ag$_2$S is further confirmed by the chiral surface Fermi arcs.
Fixing the energy at W1 and W2, we can obtain the extremely long chiral Fermi arcs in (001) surface in Fig. 3{\bf f} and 3{\bf g}, respectively. 
Owing to the large spin splitting, the spin texture should be easily observed by both ARPES and STM. 
Since two W2 Weyl points overlap with each other when they are projected into (001) surface, two Fermi arcs start from same projected Weyl point.
These extremely clear topological properties offer a good platform for both surface detection and bulk nonlinear optical transports.

\subsection*{Shift Current Analysis}
Very recently, a giant shift current was observed experimentally in WSM TaAs~\cite{BPVE_Weyl}, 
which suggests a topological effect in BPVE
and the practical applications of WSMs in optical detectors and solar energy conversion. 
Inspired by this work, 
we screened the nonlinear optical response for all the nonmagnetic WSMs in our database.
Half of them can host a large shift current above $ 100 \ \mu A/V^2$ with photon energy larger than 0.1 eV
 (Supplementary Table 2 and Supplementary Figure 1-3).
Our database shows that huge shift current can be obtained in both
infrared and visible spectrum in noncentrosymmetryic WSMs, and it provides
exotic material candidates for the study of interplay of different functions
based on WSMs.

Through perturbative analysis, the expression of shift current~\cite{secondorder} is written as
\begin{equation}
J_i(0)=2\sigma_{ijk}(0;\omega,-\omega)Re[E_j(\omega)E_k(-\omega)],
\end{equation}
where $i,j,k = x,y,z$. 
The second order optical conductivity tensor can be computed as
\begin{equation}
\sigma_{ijk}(0, \omega,-\omega)=\frac{i e^{3} \pi}{\hbar V} \sum_{\vec{k}} \sum_{m,n} f_{nm}\left(r_{mn}^{j} r_{nm}^{k;i}+r_{mn}^{k} r_{nm}^{j;i}\right) \delta\left(\hbar \omega-E_{mn}\right),
\end{equation}
where n and m are band indexes, 
$f_{nm}=f_n-f_m$ is the Fermi-Dirac distribution difference, 
$r_{nm}^{i}=A_{nm}^{i}\left(1-\delta_{n,m}\right)$ is the inter-band Berry connection,
$r_{nm}^{i;j}=\partial_{k_{j}} r_{nm}^{i}-i\left(A_{nn}^{j}-A_{mm}^{j}\right) r_{nm}^{i}$ 
depends on intermediate virtual states, 
and $E_{nm} = E_n-E_m$ is the energy difference between two bands.
The delta function $\delta\left(\hbar \omega-E_{mn}\right)$ is disposed
by a Gauss distribution with smearing factor 0.01 eV.
To avoid the numerical divergence with zero denominator, we
replace the transition-energy-related term $\frac{1}{E_{mn}}$
by $\frac{E_{mn}}{E_{mn}^{2}+\alpha^{2}}$ with $\alpha=0.04 eV$, which
is stable for the shift current with photon energy above 0.1 eV.
After a $k$-grid convergence test, we find that the change is less than 
$5\%$ with $k$-grid increasing from $100\times100\times100$ to $200\times200\times200$. 
Here, we use a grid $200\times200\times200$ in our calculations.
The calculated tensors of shift current are double-checked by the symmetry.
For convenient, we set the tensor at each $k$-point as
\begin{equation}
	\sigma^{\prime}_{ijk}(\vec{k}; 0, \omega,-\omega)=\frac{i e^{3} \pi}{\hbar V} \sum_{m,n} f_{nm}\left(r_{mn}^{j} r_{nm}^{k;i}+r_{mn}^{k} r_{nm}^{j;i}\right) \delta\left(\hbar \omega-E_{mn}\right).
\end{equation}

In the experimental measurement of WSM TaAs (ref.~\cite{BPVE_Weyl}), 
an apparent measured value $\sigma_{xxz}=34 \pm 3.7 \mu A V^{-2}$ 
with photon energy $\hbar \omega=0.117 eV$ is obtained. 
Our calculation gives a well-fitting theoretical value $\sigma_{xxz}=27 \mu A V^{-2}$ with the same photon energy.
Furthermore, ref.~\cite{BPVE_Weyl} added the reflectance of TaAs at $10.6 \mu m$ 
to the apparent measured value manually and the giant shift current 
$\sigma_{xxz}=154 \pm 17 \mu A V^{-2}$ is obtained in the end.
At the same time, $\sigma_{zzz}$ has the biggest value for TaAs in both our calculation and ref.~\cite{BPVE_Weyl}.

The largest shift current in WSMs was found
in BiTeI with space group P3, in which the component of $\sigma_{xxz}$ can
reach up to $750 \ \mu A/V^2$ at a photon 
energy 0.66 eV. It is around 5 times larger than that in TaAs. 
Moreover, four independent tensor components in BiTeI are above $200 \ \mu A/V^2$
in a large photon energy window of 0.3 to 0.7 eV (see Fig. 4{\bf a}), 
which allows large shift current in different experimental setup.
Hence the new WSM BiTeI offers a more ideal candidate
for the generation of strong BPVE. 
Considering the layered structure of BiTeI 
stacking along $z$, this component provides a natural advantage to
integrate the material into devices.

Beside the largest shift current candidate BiTeI, we also find another two kind of candidates.
One is multifunctional candidates combining topology, strong shift current, and superconductivity.
As indicated in Fig. 4{\bf b} and 4{\bf c}, the shift current tensor can reach
up to $\sigma_{zzz} = 564 \ \mu A/V^2$ and  $\sigma_{zzz} = 250 \ \mu A/V^2$ 
for the new found WSMs LaIrP and LaRhP, respectively.
In addition, since both LaIrP and LaRhP are superconductors~\cite{LaMP}
with transition temperatures $T_c = 5.3 \ K$ and $2.5 \ K$, 
these two materials will provide good platform for the study of 
the interplay among topology, superconductivity and nonlinear optical responses.
The other one is the candidates with large photon energy.
For all these WSMs with large shift current in Supplementary Table 2 and Supplementary Figure 1-3, 
we find that the peak value of $\sigma_{ijk}$ locates at an energy below 1 eV, except SrHgSn, CaBiAg and CaCdPb.
Fig. 4{\bf d} shows the shift current tensor of SrHgSn with space group $P6_3mc$ (No. 186) as an example.
One can find a peak value $\sigma_{zzz} =165 \ \mu A/V^2$ at photon energy $\sim1.8 eV$, 
which is already located at the range of red light. 

The shift current can be considered as a joint effect of the frequency-dependent 
density of states (DOS) and the wave function. 
Taking SrHgSn as an example, we can try to understand the microscopic
mechanism in these two aspects.
From the DOS of SrHgSn shown in Fig. 5{\bf b}, one can find
two peaks locating bellow and above the charge neutral point, respectively. 
The energy difference is around 1.8 eV, which is very close to the two
peaks of $\sigma_{zzz}$ in Fig. 4{\bf d}.
$\sigma^{\prime}_{zzz}(\vec{k}; 0, \omega,-\omega)$ is also calculated for each
$k$-point with the photon energy 1.8 eV and 1.95 eV near the two peaks. 
Thus we obtained that the positive contribution is mainly from the special plane 
$k_z\sim0.31$ $2\pi/c$ at photon energy around 1.8 eV, while 
$k_z\sim0.13$ $2\pi/c$ plane at photon energy around 1.95 eV has a negative contribution.
The distribution of $\sigma^{\prime}_{zzz}(\vec{k}; 0, \omega,-\omega)$
at photon energy 1.8 eV is dominated by twelve positive blue lines
caused by the excitation from band $n-1$ to bands $n+3$
and $n+4$, as shown in Fig. 5{\bf c} and 5{\bf f}. 
The distribution of $\sigma^{\prime}_{zzz}(\vec{k}; 0, \omega,-\omega)$ 
changes sharply when photon energy increasing from 1.8 eV to 1.95 eV. 
At photon energy 1.95 eV, $\sigma^{\prime}_{zzz}(\vec{k}; 0, \omega,-\omega)$ 
is dominated by two negative red rings as shown in Fig. 5{\bf d} and 5{\bf e}. 
These two red rings are derived by the light excitation from two occupied bands $n$ and $n-1$ 
to the non-occupied bands $n+7$, $n+8$, $n+9$ and $n+10$, as presented in Fig. 5{\bf g} and 5{\bf h}.
Therefore, the two peaks of $\sigma_{zzz}(0, \omega,-\omega)$ in Fig. 4{\bf d}
originate from large DOS located around $E_F-1.2 eV$ and $E_F+0.6 eV$,
and the sign changes because of the different band excitation at different $k$-points. 
One point need to emphasize is that the strong shift current at high photon energy
is not caused by the Weyl points here.

In summary, through high-throughput calculations, we have complemented the nonmagnetic topological material
databases by including the nonmagnetic WSMs.
Among them, 9 chiral structures, such as Ag$_2$S,
which allow the quantized CPGE
and have natural long surface Fermi arcs are also recognized.
Furthermore, most of them can have a large shift current,
especially for BiTeI.
Beside the two databases of nonmagnetic WSMs and their nonlinear
optical responses, the effective algorithms for searching Weyl points
by Berry phase and automatically Wannier function generation provide
powerful tools for the study of magnetic topological materials.

\section*{Methods}
The calculation starts from the experimental noncentrosymmetric crystal structures in the Inorganic Crystal Structure 
Database (ICSD, http://www2.fiz-karlsruhe.de/icsd$\_$home.html).
We take these crystal structures as the input of the full-potential local-orbital (FPLO) minimum-basis 
code~\cite{Koepernik1999}.
Here we choose the generalized gradient approximation of the Perdew-Burke-Ernzerhof
functional~\cite{perdew1996} as the exchange-correlation potential.
A dense $12\times12\times12$ k-mesh is used in the static nonmagnetic calculation. 
SOC is also included self-consistently during the workflow.
In order to get the surface states of WSMs, Green's function method~\cite{Sancho1984,Sancho1985} 
is applied with a half-infinite boundary condition.
During the DFT calculations, we assume that the ground state is nonmagnetic 
after excluding the materials with magnetic elements and large total magnetic moment.
The final database which may contains antiferromagnets is also double-checked to make sure they are indeed nonmagnetic.

\section*{data availability}
The data generated during this study are available from
the corresponding author on reasonable request.

\section*{code availability}
The codes used in this study are available from 
the corresponding author on reasonable request.

\begin{acknowledgments}
This work was financially supported by the ERC Advanced Grant No.\ 291472 `Idea Heusler', ERC Advanced Grant No.\ 742068 `TOPMAT'.
We also acknowledge funding by the DFG through SFB 1143 (project ID 247310070)
and the W$\ddot{u}$rzburg-Dresden Cluster of Excellence on Complexity and Topology
in Quantum Matter -- ct.qmat (EXC 2147, project ID 39085490).
Some of our calculations are carried on Cobra cluster of MPCDF, Max Planck society.
\end{acknowledgments}

\section*{Author contributions}
Q.X. and Y.Z. contributed equally to this work.
Q.X. collected and analysed the data and did all the calculations.
Y.Z. designed the workflow and wrote the codes for the search.
W.J. and Y.S. wrote some other supporting codes and gave technical advices.
K.K., K.B. and C.F. gave scientific advices.
Y.S. and Q.X. wrote the manuscript.
The project was supervised by C.F..

\section*{competing interests}
The authors declare no competing financial or non-financial interests.

%\nocite{*}
%\bibliographystyle{unsrt}
%\bibliographystyle{apsrev4-1}
%\bibliography{topology}% Produces the bibliography via BibTeX.

\begin{figure}[t]
\begin{center}
\includegraphics[width=0.98\textwidth]{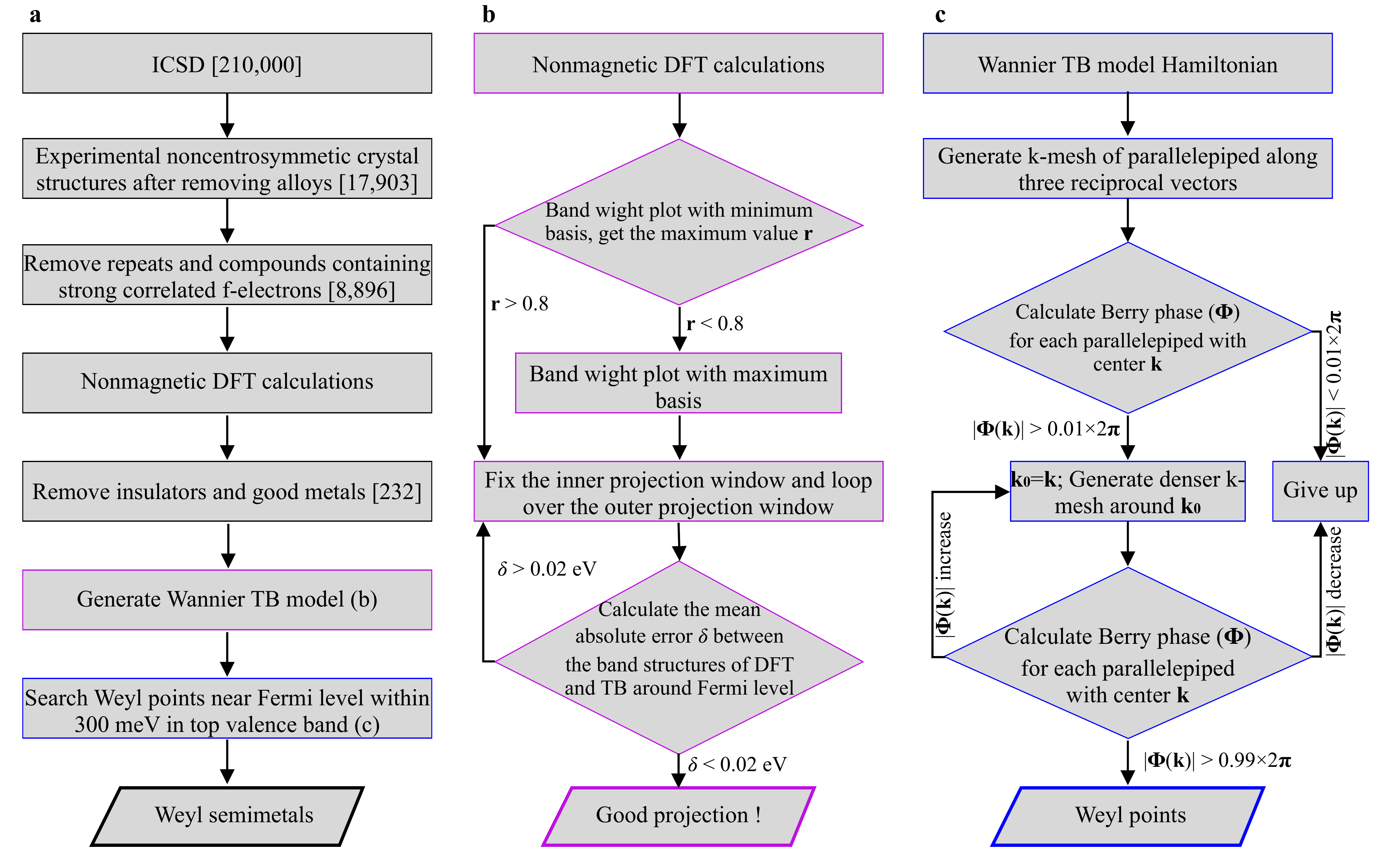}
\end{center}
\caption{
Workflow of the high-throughput calculations for the WSMs identification.
{\bf a} The experimental noncentrosymmetric crystal structures are loaded from the Inorganic Crystal Structure Database
(ICSD) after removing alloys and repeated structures.
All the compounds containing strongly correlated f-electrons are also removed from the list.
With the final input list, we perform a nonmagnetic calculation to obtain the band gap and the charge carrier density.
Here we label the compounds with magnetic elements V, Mn, Fe, Co and Ni or magnetic moments above 0.05 $\mu_{B}$ per unit cell as magnetic.  
Furthermore, for the remaining 232 semimetal phases, the Bloch wave functions are automatically projected into high-symmetry
atomic-orbital-like Wannier functions and generate the corresponding tight-binding (TB) model Hamiltonians.
Using these TB model Hamiltonians, we search the Weyl points around the Fermi energy.
{\bf b} Sub-branch of automatic Wannier function generation.
{\bf c} Sub-branch of automatic Weyl points search with the Berry phase approach method.
}
\label{band}
\end{figure}

\begin{figure}[t]
\begin{center}
\includegraphics[width=0.9\textwidth]{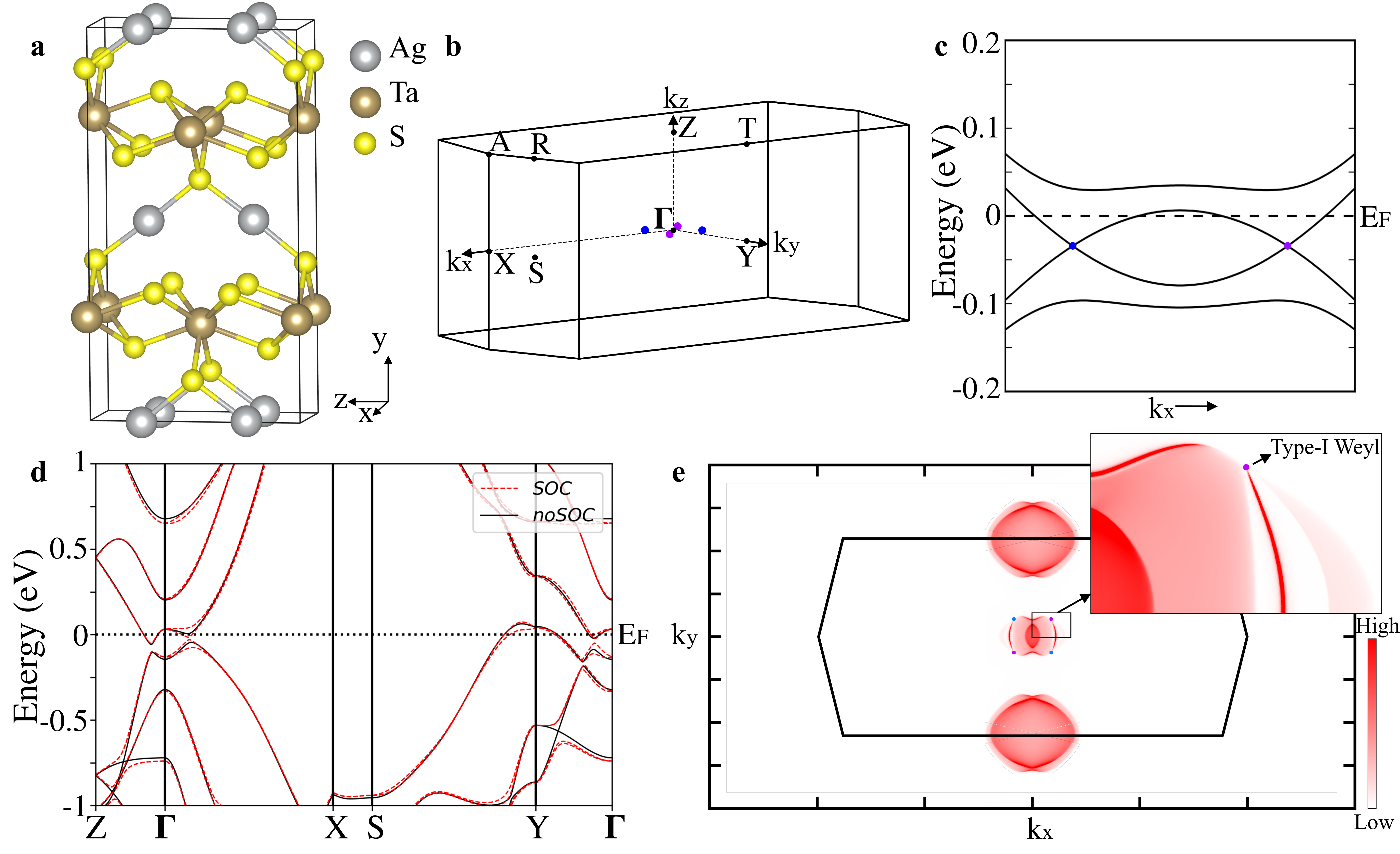}
\end{center}
\caption{
Type-I WSM TaAgS$_3$.
{\bf a} Orthorhombic crystal structure of TaAgS$_3$.
{\bf b} Brillouin zone of TaAgS$_3$ and Weyl points distribution which is obtain from Weyl point search algorithm.
{\bf c} Energy dispersion along one pair of Weyl points parallel to the $k_x$ axis.
{\bf d} Band structure with (red dot lines) and without (black solid lines) SOC along high-symmetry lines.
The band inversion appears around $\Gamma$ point.
{\bf e} Surface Fermi arcs with energy fixed at the Weyl points.
The purple and blue dots represent the Weyl points with positive 
and negative chirality, respectively.
}
\label{insulator}
\end{figure}

\begin{figure}[t]
\begin{center}
\includegraphics[width=0.9\textwidth]{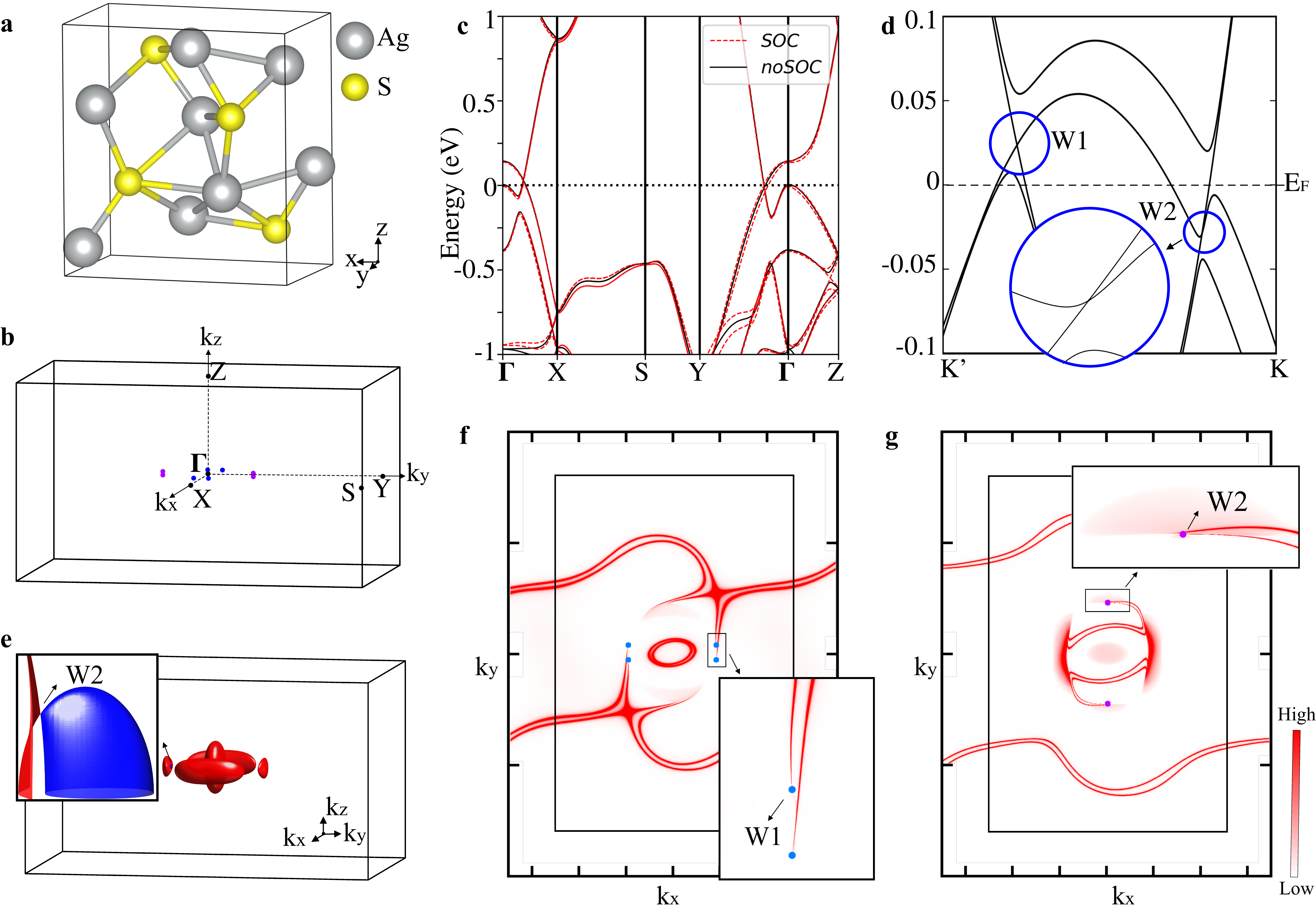}
\end{center}
\caption{
Hybrid WSM Ag$_2$S with chiral crystal symmetry.
{\bf a} Orthorhombic crystal structure of Ag$_2$S.
{\bf b} Brillouin zone and Weyl points distribution which is obtain from Weyl point search algorithm. 
{\bf c} Band structure with (red dot lines) and without (black solid lines) SOC along high-symmetry lines.
The obvious band crossing points are located at $\Gamma-X$ and $\Gamma-Y$ without SOC.
{\bf d} Energy dispersion along one pair of Weyl points with opposite chirality.
W1 and W2 represent the type-I Weyl point and the type-II Weyl point, respectively.
{\bf e} The constant energy surface in the energy of the type-II Weyl point W2. 
the linear crossing point of electron and hole pockets is the exact location of W2.
{\bf f}, {\bf g} (001) surface states with energy fixed at W1 and W2.
The purple and blue dots represent the type-I Weyl points with positive chirality 
and type-II Weyl points with negative chirality, respectively.
}
\label{metal}
\end{figure}

 \begin{figure}[t]
 \begin{center}
 \includegraphics[width=0.85\textwidth]{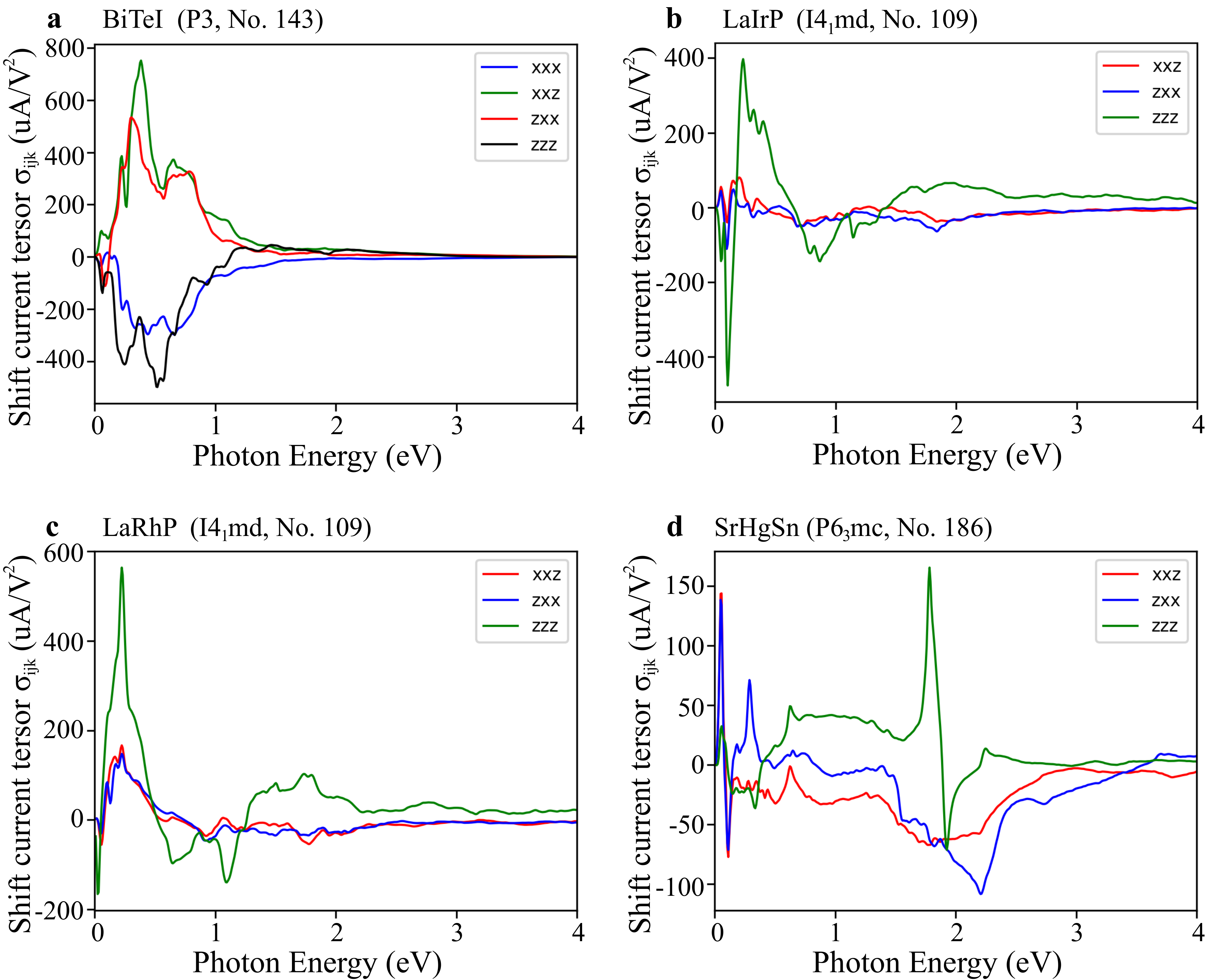}
 \end{center}
 \caption{
Shift current tensor $\sigma_{ijk}$.
It is a function of photon energy for 
{\bf a} BiTeI, {\bf b} LaIrP, {\bf c} LaRhP and {\bf d} SrHgSn.
 They are calculated based on TB model Hamiltonians generated by step {\bf b} of our algorithm.
 All of them have large shift current with tensor $\sigma_{ijk} > 100 \ \mu A/V^2$
 and huge range of photon energy which can even reach visible light ($>$ 1.65 eV).
 }
 \label{metal}
 \end{figure}
 
 \begin{figure}[t]
 \begin{center}
 \includegraphics[width=0.85\textwidth]{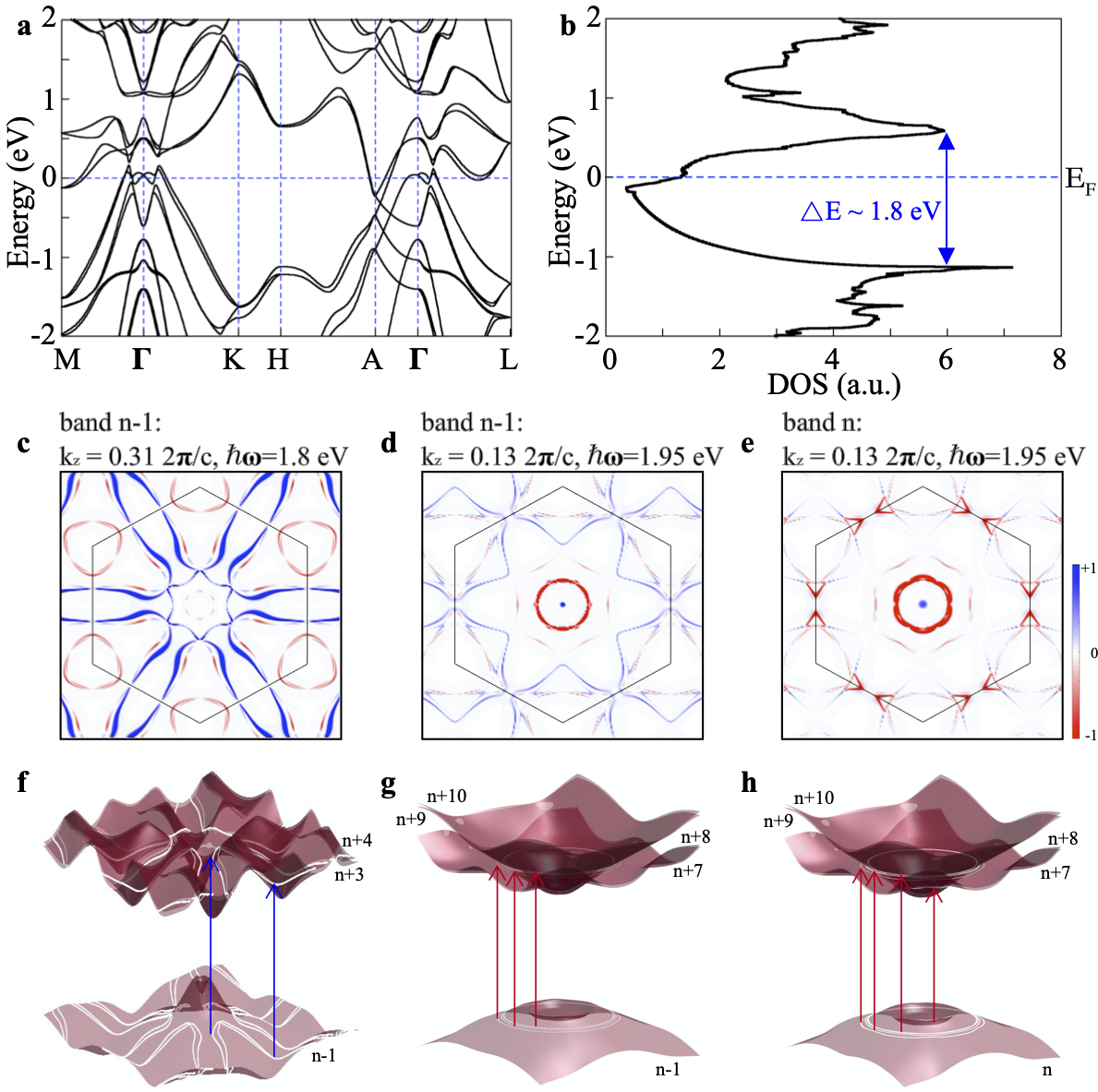}
 \end{center}
 \caption{
The mechanism analysis for $\sigma^{\prime}_{zzz}(\vec{k}; 0, \omega,-\omega)$ of SrHgSn.
{\bf a} Energy dispersion along high symmetry $k$-path. 
{\bf b} Density of states (DOS). 
{\bf c}-{\bf e} Distribution of $\sigma^{\prime}_{zzz}(\vec{k}; 0, \omega,-\omega)$ in reciprocal 
space contributed by
band $n-1$ in $k_z=0.31$ $2\pi/c$ plane with $\hbar\omega = 1.8 eV$, 
band $n-1$ in $k_z=0.13$ $2\pi/c$ plane with $\hbar\omega = 1.95 eV$, and 
band $n$ in $k_z=0.13$ $2\pi/c$ plane with $\hbar\omega = 1.95 eV$, respectively.
{\bf f}-{\bf h} Light excitation 
from band $n-1$ to bands $n+3$ and $n+4$ in $k_z=0.31$ $2\pi/c$ plane with $\hbar\omega = 1.8 eV$, 
from band $n-1$ to bands $n+7$, $n+8$, $n+9$ and $n+10$ 
in $k_z=0.13$ $2\pi/c$ plane with $\hbar\omega = 1.95 eV$, and 
from band $n$ to bands $n+7$, $n+8$, $n+9$ and $n+10$ 
in $k_z=0.13$ $2\pi/c$ plane with $\hbar\omega = 1.95 eV$, respectively.
$n$ is the number of electrons in the primitive cell. 
The blue and red arrows in {\bf f}-{\bf h} represent positive and negative contributions to $\sigma_{zzz}$.
 }
 \label{metal}
 \end{figure}
 
\end{document}